\documentclass[natbib]{emulateapj}

\input epsf
\def\s2n{S^{\prime}/N}

\submitted{}
\shorttitle{The Star Formation Rate of Supersonic MHD Turbulence}
\shortauthors{Padoan and Nordlund}

\begin{document}
\title{The Star Formation Rate of Supersonic MHD Turbulence}

\author{Paolo Padoan}
\affil{ICREA \& ICC, University of Barcelona, Marti i Franqu\`{e}s 1, E-08028 Barcelona, Spain;
ppadoan@icc.ub.edu}
\author{\AA ke Nordlund}
\affil{Niels Bohr Institute, University of Copenhagen,
Juliane Maries Vej 30, DK-2100, Copenhagen,
Denmark; aake@nbi.dk}

\begin{abstract}

This work presents a new physical model of the star formation rate (SFR),
verified with an unprecedented set of large numerical simulations of
driven, supersonic, self-gravitating, magneto-hydrodynamic (MHD) turbulence, where
collapsing cores are captured with accreting sink particles. The model depends on the relative
importance of gravitational, turbulent, magnetic, and thermal energies, expressed
through the virial parameter, $\alpha_{\rm vir}$, the rms sonic Mach number, ${\cal M}_{\rm S,0}$,
and the ratio of mean gas pressure to mean magnetic pressure, $\beta_0$. The SFR is predicted to
decrease with increasing $\alpha_{\rm vir}$ (stronger turbulence relative to gravity), to increase with 
increasing ${\cal M}_{\rm S,0}$ (for constant values of $\alpha_{\rm vir}$), 
and to depend weakly on $\beta_0$ for values typical of star forming 
regions (${\cal M}_{\rm S,0}\approx 4$-20 and $\beta_0\approx 1$-20). In the unrealistic limit of $\beta_0\to \infty$,
that is in the complete absence of a magnetic field, the SFR increases approximately by a factor of three,
which shows the importance of magnetic fields in the star formation process, even when they are
relatively weak (super-Alfv\'{e}nic turbulence). In this non-magnetized limit, our definition of the critical density
for star formation has the same dependence on $\alpha_{\rm vir}$, and almost the same dependence on
${\cal M}_{\rm S,0}$, as in the model of Krumholz and McKee, although our physical derivation does not
rely on the concepts of local turbulent pressure and sonic scale.  However, our model predicts a different
dependence of the SFR on $\alpha_{\rm vir}$ and ${\cal M}_{\rm S,0}$ than the model of Krumholz and McKee.
The star-formation simulations used to test the model result in an approximately
constant SFR, after an initial transient phase. Both the value of the SFR and its dependence on the virial parameter
found in the simulations are shown to agree very well with the theoretical predictions. A physical model of the SFR
is needed for a realistic implementation of the star formation feedback in simulations of galaxy formation,
and to retrieve the correct morphological and chemical evolution of galaxies. The new
star formation law derived in this paper is suitable for such applications.

\end{abstract}


\section{Introduction}

A physical theory of the SFR should explain why the star-formation process is slow, meaning that it converts
only a small fraction of the gas mass into stars in a free-fall time, $\tau_{\rm ff}$, both on Galactic scale
\citep{Zuckerman+Palmer74,Williams+McKee97} and on the scale of individual clouds
\citep{Krumholz+Tan07slowsf,Evans+09}. Several authors have proposed that the observed supersonic turbulence may be
responsible for keeping the SFR low by providing turbulent pressure support against the gravitational
collapse. For example, \citet{Bonazzola+87,Bonazzola+92} presented a gravitational instability analysis
that includes the effect of local turbulent pressure support; \citet{Krumholz+McKee05sfr} defined the
critical density for star formation based on the local turbulent pressure support of a Bonnor-Ebert
sphere;  \citet{Hennebelle+Chabrier08} proposed that the Salpeter stellar IMF is the result of the local
turbulent  pressure support. In all these works, the turbulent pressure is assumed to scale according to
the observed Larson velocity-size relation
\citep{Larson81,Heyer+Brunt04} or by numerical simulations.

The concept of turbulent pressure support was introduced in the context of subsonic, small-scale turbulence
by \citet{Chandrasekhar51}. It applies when the two following conditions are satisfied, $L \ll L_{\rm J}$
and $\sigma_v \ll c_{\rm S}$, where $L$ is the length scale, $L_{\rm J}$ is the Jeans length, $\sigma_v$ is
the velocity dispersion, and $c_{\rm S}$ is the sound speed. In the supersonic turbulence of star-forming
regions, both conditions are violated. As a result, the turbulence can actually trigger gravitational
collapse, causing a large-scale compression rather than preventing it. The turbulence is
responsible for much of the complex and filamentary density structure observed in molecular clouds, and
prestellar cores are likely assembled as the densest regions in this turbulent fragmentation process
\citep{Padoan+2001cores}. However, even though supersonic turbulence is able to intermittently create dense regions that are
gravitationally unstable, it does so only inefficiently, and its net effect on the large scale is that of
suppressing star formation when the total turbulent kinetic energy exceeds the total gravitational energy.

Focusing on the competition between supersonic turbulence and self-gravity, the star-formation process can
be shown to depend primarily on the ratio of the turbulent kinetic energy, $E_{\rm K},$ and the
gravitational energy, $E_{\rm G}$, of a star-forming region. This ratio may be measured by the virial parameter
introduced by \citet{Bertoldi+McKee92},
\begin{equation}
\alpha_{\rm vir} \sim {2 E_{\rm K}\over{E_{\rm G}}}= {5 \sigma^2_{\rm v,1D} R \over{G M}} ,
\label{alpha}
\end{equation}
where $\sigma_{\rm v,1D} $ is the one-dimensional rms velocity, $R$ and $M$  the cloud radius and mass
respectively, and $G$ the gravitational constant, and it has been assumed the cloud is a sphere with
uniform density. If the dynamical time is defined as the ratio of the cloud radius and the
three-dimensional rms velocity, $\tau_{\rm dyn}=R/ \sigma_{\rm v,3D}$, and using the standard definition of
the free-fall time,
$\tau_{\rm ff,0}=(3 \pi/(32 G \rho_0))^{1/2}$, the virial parameter can also be expressed as
\begin{equation}
\alpha_{\rm vir}=0.7(\tau_{\rm ff,0}/\tau_{\rm dyn})^2.
\label{alpha_timescales}
\end{equation}

\citet{Krumholz+McKee05sfr} derived a theoretical model where the SFR is primarily controlled by the virial
parameter. In this model, it is assumed that the gas mass above some critical density, $\rho_{\rm cr}$, is
gravitationally unstable, and the fraction of this unstable mass is computed assuming the gas density obeys
a Log-Normal pdf \citep{Nordlund+Padoan_Puebla98}. Following \citet{Padoan95}, the critical density is
defined through the comparison of the Jeans' length and the sonic-scale, $\lambda_{\rm s}$, which is the
scale where the turbulent velocity differences are of the order of the speed of sound. The critical density is
equivalent to that of the critical Bonnor-Ebert mass of size $\lambda_{\rm s}$. The idea of relying
on the density pdf was also exploited in \citet{Padoan+Nordlund02imf,Padoan+Nordlund04bd} to explain
the stellar IMF and the origin of brown dwarfs, and by \citet{Padoan95} to model the SFR.

The model of \citet{Krumholz+McKee05sfr} was calibrated and tested using low-resolution SPH simulations by
\citet{Vazquez+03sfr_sonic}. Because of the important role of turbulent energy in this model, low-resolution
simulations are inadequate. They do not develop an inertial range of turbulence and are
expected to produce a too large SFR -- which they do, as recognized in a later paper, based on
higher-resolution grid simulations, by some of the same authors \citep{Vazquez+05sfr_mhd}. However,
\citet{Krumholz+McKee05sfr} estimated a rather low SFR from the simulations of \citet{Vazquez+03sfr_sonic}
by fitting only their early evolution. We argue this is a transient phase of accelerated SFR and should not be
used to test the model (although it cannot be excluded that real molecular clouds experience such a phase of
accelerated star formation, the initial transient phase in the simulations is of numerical origin, due to the sudden
inclusion of self-gravity in our case, or to the memory of artificial initial conditions in \citet{Vazquez+03sfr_sonic},
where gravity is included from the beginning). A new set of larger simulations is needed to properly test the
theoretical model. Because the model by \citet{Krumholz+McKee05sfr} does not include the effect of magnetic
fields, a new model based on magneto-hydrodynamic (MHD) turbulence should be derived, and this new MHD
model should be tested with large numerical simulations as well.

In this work we propose such a new MHD model of the SFR, and we test it with an unprecedented set of large numerical
simulations of driven, supersonic, self-gravitating, MHD turbulence, where collapsing cores are represented by accreting sink
particles.  Both the model and the simulations are limited to the case of an isothermal gas, and the effect of deviations from
the isothermal behavior are not addressed. To model the process of star formation we must include gravitational, turbulent,
magnetic, and thermal energies. Here we express their relative importance through the virial parameter, $\alpha_{\rm vir}$,
the rms sonic Mach number, ${\cal M}_{\rm S,0}$, and the mean gas pressure to mean magnetic pressure, $\beta_0$, and we
derive a model that depends explicitly on all three non-dimensional parameters. In the non-magnetized limit of $\beta_0\to \infty$,
our definition of the critical density for star formation has the same dependence on $\alpha_{\rm vir}$ and ${\cal M}_{\rm S,0}$
as in the model of \citet{Krumholz+McKee05sfr}, but our derivation does not rely on the concepts of local turbulent pressure
support and sonic scale as in that work. However, even in this non-magnetized case, our model predicts a different dependence
of the SFR on $\alpha_{\rm vir}$ and ${\cal M}_{\rm S,0}$ than the model of \citet{Krumholz+McKee05sfr}, because we assume
that regions exceeding the critical density turn into stars on a timescale given by their local free-fall time, rather than
the free-fall time of the mean density.

Although the non-magnetized case can be derived from our MHD model in the limit of $\beta_0\to \infty$, we structure
the paper by first deriving the critical density in the purely hydrodynamic (HD) case (\S2), and then in the general MHD case
(\S3). Likewise, in \S4 we first present a simple model for the density pdf in the HD case, and then generalize
the approach to MHD turbulence. In \S5 we derive the model predictions for the SFR and in \S6 we present our
numerical simulations of HD and MHD self-gravitating turbulence. The comparison between the model and the simulations
is presented in \S7, results are discussed in \S8, and conclusions are summarized in \S9.

\section{Critical density in HD Turbulence}

In the hydrodynamic (HD) case, the main source of pressure in the postshock gas is the thermal pressure, so
the shock jump conditions are given by the balance of thermal pressure and ram pressure:
\begin{equation}
\rho_{\rm HD} \,c_{\rm S}^2 = \rho_0 (v_0/2)^2 ,
\label{eq1}
\end{equation}
where $c_{\rm S}$ is the sound speed, $\rho_0$ and $\rho_{\rm HD}$ the preshock and postshock gas
densities, and $v_0/2$ the shock velocity. Because we use this equation to estimate a characteristic postshock
density in the HD case, $\rho_{\rm HD}$, we choose the mean gas density, $\rho_0$, as the preshock
density, and half the rms velocity, $v_0$, as the shock velocity. Assuming an ensemble of eddies with a
randomly oriented velocity of mean magnitude $v_0$, the average collision velocity is also $v_0$. However, the shock
velocity is half of that average collision velocity because the postshock layer is confined by two shocks,
each with velocity $v_0/2$. The characteristic density is then given by:
\begin{equation}
\rho_{\rm HD} = \rho_0 {\cal M}_{\rm S,0}^2/4 ,
\label{eq2}
\end{equation}
where ${\cal M}_{\rm S,0}$ is the rms sonic Mach number, and the characteristic thickness, $\lambda_{\rm
HD}$, of the postshock layers is:
\begin{equation}
\lambda_{\rm HD} = (\theta\, L_0) \, 4 / {\cal M}_{\rm S,0}^2 ,
\label{eq3}
\end{equation}
where $L_0$ is the size (e.g. the diameter for a sphere) of the system and $\theta \, L_0$, with $\theta \le 1$,
is the turbulence integral scale. Because the turbulence velocity scaling is approximately $v\propto \ell^{1/2}$, this
characteristic thickness is practically scale-independent (it would have been the same if derived at any
other scale, not only at the integral scale). The local condition for collapse is that $M_{\rm HD}(\rho) \ge M_{\rm BE}(\rho)$,
where $M_{\rm BE}$ is the Bonnor-Ebert mass \citep{Bonnor56,Ebert57} with external density equal to the postshock
density $\rho$,
\begin{equation}
M_{\rm BE} = 1.182 \, c_{\rm S}^3\, /(G^{3/2}\rho^{1/2})
\label{eq_mbe}
\end{equation}
and $M_{\rm HD}$ is the mass of a uniform sphere of radius $\lambda_{\rm HD}/2$,
$M_{\rm HD}(\rho)=(4/3)\pi (\lambda_{\rm HD}/2)^3 \, \rho$.

Because the thickness, $\lambda_{\rm HD}$, is scale independent, the condition for collapse
can be used to define a scale-independent critical density for collapse. The local density depends on the
distribution of local shock velocity and preshock density and is known to follow a Log-Normal pdf
\citep{Vazquez-Semadeni94,Padoan+97imf,Nordlund+Padoan_Puebla98,Ostriker+2001,
Li+04,Kritsuk+07,Beetz+08,Lemaster+Stone08,Federrath+08}, so there is a finite probability that a
region exceeds the critical density and undergoes collapse. We therefore define the critical density for
star formation, $\rho_{\rm cr,HD}$, as the minimum density that satisfies the local condition for collapse:
\begin{equation}
M_{\rm HD}(\rho_{\rm cr,HD}) = M_{\rm BE} (\rho_{\rm cr,HD}).
\label{eq_condition_hd}
\end{equation}
which yields:
\begin{equation}
\rho_{\rm cr,HD}/ \rho_0 = 0.067 \, \theta^{-2} \alpha_{\rm vir} \, {\cal M}_{\rm S,0}^2 ,
\label{eq_rhocr_hd}
\end{equation}
where $\alpha_{\rm vir}$ is the virial parameter defined above in equation (\ref{alpha}), and can be
re-written as
\begin{equation}
\alpha_{\rm vir} = 5 v_0^2 / (\pi G \rho_0 L_0^2) ,
\label{eq_alpha_b}
\end{equation}
assuming the system is a uniform sphere of radius $L_0/2$, mean gas density $\rho_0$, and three-dimensional
rms turbulent velocity $v_0$.

The critical density defined by equation (\ref{eq_rhocr_hd}) has the same dependence on $\alpha_{\rm vir}$
and almost the same dependence on ${\cal M}_{\rm S,0}$ as the critical density derived by \citet{Krumholz+McKee05sfr}.
However, the critical density has been derived here without any reference to the concepts of turbulent pressure support
and sonic scale. On the contrary, our derivation is based on the idea that the turbulence is a trigger of local gravitational
instabilities through its dynamical pressure. This physical difference between the two derivations is reflected by the
Mach number dependence. In \citet{Krumholz+McKee05sfr}, $\rho_{\rm cr,HD} \sim {\cal M}_{\rm S,0}^{2/p - 2}$,
where $p$ is the exponent of the velocity-size relation, $v\propto \ell^p$, which gives the same dependence on Mach
number as in our model,  $\rho_{\rm cr,HD} \sim {\cal M}_{\rm S,0}^2$, only for the specific value of $p=1/2$.
In supersonic turbulence, however, the scaling exponent is not necessarily identical to the Burgers value of $p=1/2$.
Numerical simulations yield somewhat smaller values based on the second order velocity structure functions,
or somewhat larger ones based on the first order \citep{Kritsuk+07}.

The ratio between the characteristic density, $\rho_{\rm HD}$, and the critical density, $\rho_{\rm cr,HD}$, is
independent of ${\cal M}_{\rm S,0}$,
\begin{equation}
\rho_{\rm HD}/ \rho_{\rm cr,HD} = 3.521\, \theta^{2} \alpha_{\rm vir}^{-1},
\label{eq6b}
\end{equation}
which anticipates the result that the mass fraction with density above $\rho_{\rm cr,HD}$ (and hence
the SFR), must have a rather weak Mach number dependence (despite the strong dependence of $\rho_{\rm cr,HD}$
on ${\cal M}_{\rm S,0}$), and must increase with decreasing $\alpha_{\rm vir}$ (weaker turbulence relative to gravity).

In numerical simulations, the integral scale of the turbulence is somewhat smaller than the system size ($\theta<1$).
For example, in our simulations of supersonic turbulence driven in the range of wavenumbers $1\le k \le2$ ($k=1$ corresponds to
the box size), $\theta \approx 0.35$ (including a correction factor discussed in \citet{Wang+George02}). We adopt this value of
$\theta$ when we compare the models with the simulations in \S7. If star-forming regions are driven on very large scales,
for example by the expansion of supernova remnants
\citep{Korpi+99,Kim+01sn,deAvillez+Breitschwerdt05,Joung+MacLow06sn,Tamburro+09},
the turbulence integral scale could be much larger than the size of individual star-forming regions. However, in our model
$\theta L_0$ is the characteristic scale of regions of compression with velocity of order the flow rms velocity, $v_0$, with
$v_0$ measured within the region of size $L_0$. We therefore adopt the same value of $\theta=0.35$ as
estimated in the simulations.  With $\theta=0.35$, the critical number density is
\begin{equation}
n_{\rm cr,HD}/ n_0 = 0.547\, \alpha_{\rm vir}\, {\cal M}_{\rm S,0}^2 .
\label{eq7}
\end{equation}

Adopting characteristic parameters of molecular clouds on a scale of 10~pc, $\alpha_{\rm vir}\approx 1.6$, $n_0\approx
200$~cm$^{-3}$, and ${\cal M}_{\rm S,0}\approx 20$, we get a characteristic number density of $n_{\rm
HD}\approx 2.0\times 10^4$~cm$^{-3}$ from equation (\ref{eq2}), reasonable for prestellar cores, and a factor of 3.5
below the critical number density, $n_{\rm cr,HD}\approx 350.1\, n_0 \approx 7.0 \times 10^4$~cm$^{-3}$.
The critical overdensity factor of 350.1 is somewhat larger than the value of 275 derived from equation (27)
of \citet{Krumholz+McKee05sfr}, using the same values of $\alpha_{\rm vir}$ and ${\cal M}_{\rm S,0}$ (notice that their
Mach number is 1D, so a factor of $3^{1/2}$ smaller than ours) and assuming $\phi_x=1.12$ for their numerical coefficient
(their best fit to numerical simulations).

\section{Critical density in MHD Turbulence}

We now consider the magneto-hydrodynamic (MHD) case. Including both thermal and magnetic pressures, and using
$v_0/2$ for the shock velocity, like in equation (\ref{eq1}), the pressure balance condition for MHD shocks is:
\begin{equation}
\rho_{\rm MHD} (c_{\rm S}^2+v_{\rm A}^2/2 )= \rho_0 (v_0/2)^2 ,
\label{eq_pressure}
\end{equation}
where $v_{\rm A}$ is the Alfv\'{e}n velocity in the postshock gas defined by the postshock magnetic field
perpendicular to the direction of compression. Because the field is amplified only in the direction perpendicular to the
compression, the postshock perpendicular field is comparable to the total postshock field\footnote{
For the magnetic field strength in the postshock gas we can write $B^2_{\perp}=B^2-B^2_{\parallel}=B^2-B^2_{0,\parallel}$,
where the second equality is from the fact that the component parallel to the direction of the compression is not amplified.
If we take an average, assuming a random orientation of the magnetic field relative to the direction of compression, we get
$\langle B^2_{\perp}\rangle=\langle B^2\rangle - B_0^3/3$, and hence
$\langle B^2_{\perp}\rangle / \langle B^2\rangle=1-(B^2_0/\langle B^2\rangle)/3$. Thus, on the average, the relative error
in $\beta$ as a result of assuming $B=B_{\perp}$ is $(B^2_0/\langle B^2\rangle)/3$, which is typically of order 1\% or less.
}, and we can write,
$v_{\rm A}\approx B/(4\, \pi \rho)^{1/2}$, where $B$ is the postshock magnetic field and $\rho$ the
postshock gas density. The characteristic gas density and thickness of postshock layers are thus given by:
\begin{equation}
\rho_{\rm MHD} = \rho_0  ({\cal M}_{\rm S,0}^2 / 4)  \left(1+\beta^{-1}\right)^{-1} ,
\label{eq_rhomhd}
\end{equation}
\begin{figure}[t]
\includegraphics[width=\columnwidth]{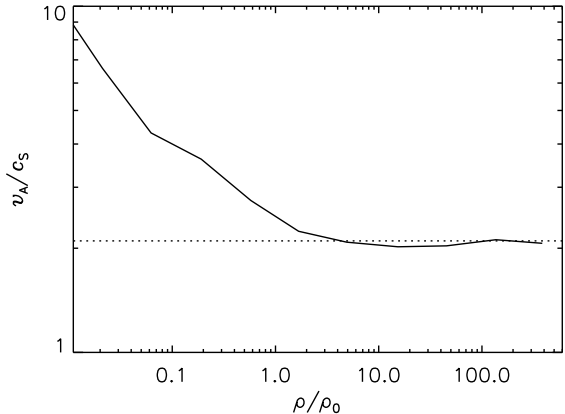}
\caption[]{Mean Alfv\'{e}n velocity, in units of the sound speed, versus gas density, in units of the mean
density, for a snapshot of the $1,000^3$ MHD turbulence simulation prior to the inclusion of self-gravity.
The Alfv\'{e}n velocity is essentially independent of density for densities above the mean. The
dotted line shows the mean value of 0.21, computed for densities larger than twice the mean.}
\label{f1}
\end{figure}
\begin{equation}
\lambda_{\rm MHD} = (\theta\, L_0) ({\cal M}_{\rm S,0}^2 / 4)^{-1}  \left(1+\beta^{-1}\right) ,
\label{eq_lambdamhd}
\end{equation}
where we have introduced the ratio of gas to magnetic pressure in the postshock gas,
$\beta=2\, c_{\rm S}^2 / v_{\rm A}^2$. In the limit of $\beta \to \infty$, these expressions reduce to the
corresponding HD ones, given by equations (\ref{eq2}) and (\ref{eq3}).
The value of $\lambda_{\rm MHD}$ is not scale independent. Its scale dependence is at the heart
of the relation between the exponent of the Salpeter stellar IMF and the turbulent velocity power spectrum,
in the IMF model of Padoan and Nordlund (2002). However, we can still define a characteristic thickness, and
hence a characteristic critical density, as in the HD case, because the average postshock Alfv\'{e}n velocity,
$v_{\rm A}$ (and the corresponding postshock $\beta$), is only very weakly dependent on density. In numerical
simulations of supersonic and super-Alfv\'{e}nic turbulence, it is found that, although $v_{\rm A}$ has a very large
scatter for any given density, its mean value is nearly density independent, corresponding to a mean relation
approaching $B\propto \rho^{1/2}$ for a very weak mean magnetic field \citep{Padoan+Nordlund99MHD}. In the
specific MHD simulation used in this work, the mean value of $v_{\rm A}$ is almost exactly constant for any density
$\rho \gtrsim 2 \rho_0$ (see Figure~\ref{f1}). Zeeman splitting measurements of the magnetic field strength in
molecular cloud cores are also consistent with an average value of $v_{\rm A}$ nearly independent of density
\citep{Crutcher99}.

Like in the HD case, we define the critical density as the density above which a uniform sphere of radius
$\lambda_{\rm MHD}/2$ is gravitationally unstable, assuming that variations in the thickness around $\lambda_{\rm MHD}$
are not strongly correlated with the density variations. To account for both thermal and magnetic support, we
adopt the approximation of the critical mass for collapse, $M_{\rm cr}$, introduced by \cite{McKee89},
\begin{equation}
M_{\rm cr} \approx M_{\rm BE} + M_{\phi},
\label{eq_mcr}
\end{equation}
where $M_{\phi}$ is the magnetic critical mass for a sphere of radius $R$, mean density equal to the postshock
density $\rho$, and constant mass-to-flux ratio,
\begin{equation}
M_{\rm \phi} = 0.17 \pi R^2 B / G^{1/2} = 0.387 v_{\rm A}^3/(G^{3/2}\rho^{1/2})
\label{eq_phi}
\end{equation}
where the numerical coefficient 0.17 is from \citet{Tomisaka+88} (see also \citet{Nakano+Nakamura78} for the
case of an infinite sheet, and \citet{McKee+Ostriker07} for a discussion of ellipsoidal clouds and other geometries).
The critical density is defined by the condition,
\begin{equation}
M_{\rm MHD}(\rho_{\rm cr,MHD}) = M_{\rm BE} (\rho_{\rm cr,MHD})+ M_{\phi}(\rho_{\rm cr,MHD}),
\label{eq_condition}
\end{equation}
where $M_{\rm MHD}(\rho)=(4/3)\pi (\lambda_{\rm MHD}/2)^3 \rho$. Equation (\ref{eq_condition}) results in the
following expression for the critical density as a function of the three non-dimensional parameters,
$\alpha_{\rm vir}$, ${\cal M}_{\rm S,0}$, and $\beta$:
\begin{equation}
     {\rho_{\rm cr,MHD}\over \rho_0} = 0.067 \, \theta^{-2} \alpha_{\rm vir} \, {\cal M}_{\rm S,0}^2
     {(1+0.925\beta^{-{3\over 2}})^{2\over 3} \over (1+\beta^{-1})^2},
\label{eq_rhocrmhd}
\end{equation}
which is smaller than $\rho_{\rm cr,HD}$ for any value of $\beta$, and reduces to the expression for $\rho_{\rm cr,HD}$
given by equation (\ref{eq_rhocr_hd}), in the limit of $\beta \to \infty$. The relative ratio of characteristic to critical density in
MHD and HD is given by the following function of $\beta$:
\begin{equation}
{\rho_{\rm MHD} / \rho_{\rm cr,MHD}  \over  \rho_{\rm HD}/ \rho_{\rm cr,HD}} = {(1+\beta^{-1})  \over (1+0.925\beta^{-{3\over 2}})^{2\over 3}},
\label{eq_ratio}
\end{equation}
This ratio is slightly larger than unity for any value of $\beta$ (with a maximum of $\approx 1.3$ at $\beta\approx 0.86$),
suggesting that star formation should be slightly more likely in MHD turbulence than in the HD case. However, due to the
less broad gas density pdf in the MHD case (see the next section), the net result is instead a lower SFR in MHD than in HD.

We have verified that the average value of $\beta$ is nearly
independent of density in the MHD simulation used to generate the initial condition for the MHD
star-formation simulations described in \S6. In that simulation, the rms sonic Mach number is
${\cal M}_{\rm S,0}\approx 9$ and the mean Alfv\'{e}n
velocity $v_{\rm A,0}=0.3 \, c_{\rm S}$, computed with the mean density and mean magnetic field. However,
the rms magnetic field is amplified by the turbulence, so the actual Alfv\'{e}n velocity should be computed
as the {\it local} absolute value of $B$ divided by the {\it local} value of the density, which gives,
$v_{\rm A}=\langle |B|/(4\pi \rho)^{1/2}\rangle=2.1\, c_{\rm S}$, if averaged over all regions with density larger than
twice the mean (the Alfv\'{e}n velocity introduced in eq. (\ref{eq_pressure}) is measured in the postshock gas, so
it should be estimated as an average in over-dense regions). Figure~\ref{f1} shows the mean Alfv\'{e}n
velocity as a function of the gas density in the snapshot used as the initial condition for the MHD
star-formation simulations (see \S~6). The Alfv\'{e}n velocity is almost exactly constant at densities
above the mean. 

In numerical simulations of super-Alfv\'{e}nic turbulence, the rms magnetic field is the result of the
amplification of some weak initial field by compressions and, possibly,
by a turbulent dynamo. These simulations typically start from an initially uniform field, $B_0$, which is
also the conserved mean magnetic field. It would be useful to relate our postshock $\beta$ to the ratio of gas to
magnetic pressure computed with the mean magnetic field, $B_0$, and the mean gas density, $\rho_0$,
$\beta_0=2\, c_{\rm S}^2/v_{\rm A,0}^2$, where $v_{\rm A,0}^2=B_0^2/(4\pi\rho_0)$. An approximate relation 
for the dependence of $\beta$ on $\beta_0$ and ${\cal M}_{\rm S,0}$ can be derived based
on flux freezing, on the simplified MHD shock jump conditions without thermal pressure (where we assume the
characteristic shock velocity is $v_0/2$, as in equations (\ref{eq1}) and (\ref{eq_pressure})), and neglecting dynamical
alignement of flow velocity and magnetic field:
\begin{equation}
\beta\approx b \, \beta_0^{1/2}{\cal M}_{\rm S,0}^{-1},
\label{eq_betamodel}
\end{equation}
With the MHD simulation of this work, we derive $b=0.22$ when $\beta$ is computed from the mean squared value of
$v_{\rm A}$ averaged over the whole computational box (not limited to over-dense regions). We find that equation
(\ref{eq_betamodel}) is a very good approximation also for the three $1,024^3$ simulations of \citet{Kritsuk+09}, where 
${\cal M}_{\rm S,0}\approx 10$, and $\beta_0=0.2$, 2, and 20 (it overestimates $\beta$ by approximately 20\% for 
$\beta_0=0.2$ and 2.0, and underestimates it by approximately 2\% for $\beta_0=20$). However, if $\beta$ is computed
from the mean squared $v_{\rm A}$ averaged above a certain density, we find that, as we increase the value of that 
density threshold, the value of $\beta$ becomes gradually independent of $\beta_0$. For densities larger than 50 times 
the mean, for example, all three simulations yields $\beta \approx 1$. This can be understood as due to the tendency 
of the strongest density enhancements to originate from compressions along the magnetic field direction. This tendency
becomes stronger for decreasing values of  ${\cal M}_{\rm A,0}$, or, at constant ${\cal M}_{\rm S,0}$, for decreasing values
of $\beta_0$, as documented by the increased alignement of flow velocity and magnetic field \citep{Kritsuk+09_self_org}.

Given the difficulty of deriving a robust value of the effective postshock $\beta$ to be used in the model for the critical density, 
we estimate $\beta$ based on the density pdf, as explained in the next section. We will derive a value of $\beta=0.39$.
The compilations of OH and CN Zeeman measurements by \citet{Troland+Crutcher08} 
and \citet{Falgarone+08} give an average value of $\beta=0.34$ and 0.28 respectively, using the 
line-of-sight magnetic field strength of their 3-$\sigma$ detections, assuming a temperature $T=10$ and 50~K respectively,
and averaging the values of $\beta$ of the individual cores. These estimated values of $\beta$ are very close to that 
derived in the next section based on the density pdf.

\section{Gas Density PDF}

We can estimate the gas mass fraction that is turned into stars by computing the mass fraction above the critical density, as in
\citet{Krumholz+McKee05sfr}. For given values of $\alpha_{\rm vir}$, ${\cal M}_{\rm S,0}$, and $\beta$ (or $\beta_0$), the critical
density is fixed, and the mass fraction above the critical density is determined by the density pdf. In the HD case, the density pdf is
known to be Log-Normal, with a standard deviation depending on the rms Mach number. Following the numerical results of
\cite{Padoan+97imf} for the Mach number dependence, the pdf is given by:
\begin{equation}
p_{\rm HD}(x)dx=\frac{ x^{-1} } { (2\pi\sigma_{\rm HD}^2)^{1/2}}\, {\rm exp} \left[ -\frac{({\rm ln}x + \sigma_{\rm HD}^2/2)^2}{2\, \sigma_{\rm HD}^2} \right] dx 
\label{eq15} 
\vspace{0.1cm}
\end{equation}
and the standard deviation, $\sigma$, is given by
\begin{equation}
\sigma_{\rm HD}^2 \approx  {\rm ln}\left[1+\left(\frac{{\cal M}_{\rm S,0}}{2}\right)^2 \right]
\label{eq17}
\end{equation}
Equation (\ref{eq17}) for the standard deviation of the logarithm of the overdensity, ${\rm ln}x$,
implies a simple expression for the standard deviation, $\sigma_{x,{\rm HD}}$, of the overdensity, $x$,
\begin{equation}
\sigma_{x,{\rm HD}}\approx {\cal M}_{\rm S,0} / 2
\label{eq18}
\end{equation}

In the MHD case the density pdf may deviate from the Log-Normal and it may depend on both the sonic and the
Alfv\'{e}nic Mach numbers. \citet{Lemaster+Stone08} have shown that the density pdf in supersonic MHD
simulations with a strong field, corresponding to a mean value of $\beta_0=0.02$,  is very similar to the
density pdf in the HD case. Assuming a Log-Normal pdf, the averaged results given in their Table~1 correspond 
to the relation $\sigma_{x,{\rm MHD}}\approx C \, {\cal M}_{\rm S,0}/2$, with $C\approx 0.8$ at ${\cal M}_{\rm S,0}<4$, 
and $C$ decreasing with increasing Mach number for ${\cal M}_{\rm S,0}>4$. In their largest Mach number run they
find $C\approx 0.66$ with ${\cal M}_{\rm S,0}\approx 6.7$, not far from the value of $C\approx 0.53$ derived below 
(see equation (\ref{eq21b})) from our MHD run with an even larger Mach number, ${\cal M}_{\rm S,0}\approx 9$.
In the absence of a detailed numerical study, including different values of $\beta_0$ and large values of ${\cal M}_{\rm S,0}$, 
here we derive a simple model for the density pdf in the MHD case, based on arguments inspired by the HD case.
We assume that the pdf can be approximated by a Log-Normal also in the MHD case,
\begin{equation}
p_{\rm MHD}(x)dx=\frac{ x^{-1} } { (2\pi\sigma_{\rm MHD}^2)^{1/2}}\, {\rm exp} \left[ -\frac{({\rm ln}x + \sigma_{\rm MHD}^2/2)^2}{2\, \sigma_{\rm MHD}^2} \right] dx
\label{eq_pdf_mhd}
\vspace{0.2cm}
\end{equation}
at least in the
super-Alfv\'{e}nic regime that we think is relevant for molecular clouds (Padoan and Nordlund 1999;
Lunttila et al. 2008,2009). This may not be a good approximation for the low density tail of the pdf, but
for the present purpose we are primarily interested in the high density tail. To derive an expression for
$\sigma_{\rm MHD}$, we first show that the dependence of $\sigma_{\rm HD}$ on ${\cal M}_{\rm S,0}$
can be obtained with a simple derivation, and we then apply the same derivation to the MHD case.

Let's consider a cubic box of size $L_0$ swept by a single compression of sonic Mach number ${\cal M}_{\rm
S,0}$ in one direction and therefore accumulating all the mass in a postshock layer of size $L_0$ and
density and thickness given by equations (\ref{eq2}) and (\ref{eq3}) respectively, with
$\theta=1$. The standard deviation of the density, $\sigma_{\rho}$, is given by
\begin{equation}
\sigma_{\rho}^2= \frac{1}{V}\int_V (\rho-\rho_0)^2dV
\label{eq19}
\end{equation}
where $V$ is the volume, and the integral is over the whole volume. In our simple model, the density is
either zero outside of the layer, or $\rho =\rho_{\rm HD}\gg \rho_0$ inside the layer. The integral is
therefore approximately equal to $\rho_{\rm HD}^2$ times the volume of the layer, $V_{\rm layer}$:
\begin{equation}
\sigma_{\rho}^2 \approx \frac{1}{V}  (\rho_{\rm HD}^2 V_{\rm layer}) = \frac{\lambda_{\rm HD} L_0^2}{L_0^3}
\rho_{\rm HD}^2=\rho_0^2  {\cal M}_{\rm S,0}^2/4
\label{eq20}
\end{equation}
where we have used equations (\ref{eq2}) and (\ref{eq3}) in the last equality. This result is equivalent to
equation (\ref{eq18}) that was derived from numerical simulations
of supersonic turbulence \citep{Padoan+97imf,Nordlund+Padoan_Puebla98}, and was recently confirmed
by \cite{Brunt+10pdf}, based on extinction maps of the Taurus molecular cloud. Following the same derivation
in the MHD case we obtain:
\begin{equation}
\sigma_{x,{\rm MHD}}\approx (1+\beta^{-1})^{-1/2} {\cal M}_{\rm S,0} / 2,
\label{eq21}
\end{equation}
\begin{figure}[t]
\includegraphics[width=\columnwidth]{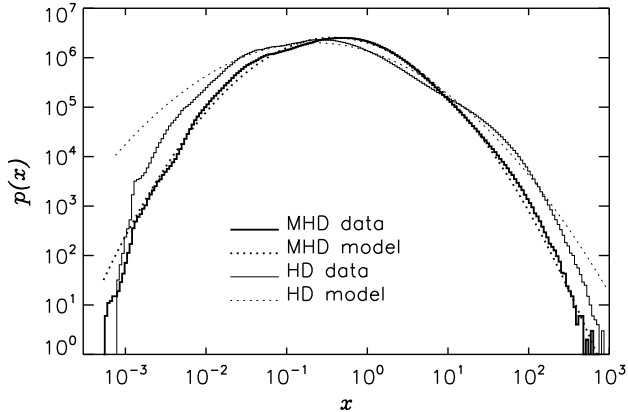}
\caption[]{Pdf of gas density for the MHD and HD snapshots used as initial conditions for the
star-formation simulations (solid lines). The Log-Normal models used in this work are also shown (dotted lines).  \\}
\label{f2}
\end{figure}
corresponding to
\begin{equation}
\sigma_{\rm MHD}^2\approx {\rm ln}\left[  1 +  \left(\frac{{\cal M}_{\rm S,0}}{2}\right)^2 (1+\beta^{-1})^{-1} \right].
\label{eq_sigma_mhd}
\end{equation}

As explained at the end of the previous section, we cannot rely on equation (\ref{eq_betamodel}) to derive the
effective postshock $\beta$ from the values of $\beta_0$ and ${\cal M}_{\rm S,0}$. This is further illustrated by
the fact that the density pdfs in the three $1,024^3$ simulations of \citet{Kritsuk+09} (where 
${\cal M}_{\rm S,0}\approx 10$ and $\beta_0=0.2$, 2, and 20) are almost indistinguishable from each other,
for densities above the peak of the pdfs. This is again interpreted as due to the growing tendency
of large density enhancements to result from compressions parallel to the magnetic field, as the mean magnetic 
field strength increases, or the value of $\beta_0$ decreases. Based on our model for the standard deviation of the pdf,
equation (\ref{eq21}), the fact that the pdf does not change with $\beta_0$ implies that $\beta$ is independent of $\beta_0$.
We fit this simple pdf model to the actual density pdf of our MHD run to derive the corresponding $\beta$. The
functional form of equation (\ref{eq_sigma_mhd}) should be confirmed by numerical simulations. However, this equation
has been derived under the same assumptions, and using the same meaning of the postshock $\beta$, as in the derivation
of the critical density. This justifies our approach of deriving $\beta$ by fitting the density pdf.

Figure~\ref{f2} compares the HD and MHD model pdfs to the actual pdfs of the snapshots used as initial conditions 
for the star-formation simulations. The MHD model provides an excellent fit to the high density tail of the pdf, for over 5 
orders of magnitude in probability. At the highest densities, the HD model predicts a slightly larger probability than in 
the HD simulation, a discrepancy that may be attributed to the limited numerical resolution,
and would likely be reduced if the numerical pdf were the result of a time average of many snapshots (which would
also improve the fit of the low density tail of the pdf). The best fit to the MHD pdf is obtained with $\beta=0.39$. 
We therefore adopt this value as the postshock $\beta$ of our model for ${\cal M}_{\rm S,0}\approx 10$ and
$0.2 \lesssim \beta_0 \lesssim 20$. 

Based on equation (\ref{eq21}), this value of $\beta$ gives
\begin{equation}
\sigma_{x,{\rm MHD}}\approx 0.53 \, {\cal M}_{\rm S,0} / 2 \approx 0.53 \, \sigma_{x,{\rm HD}}
\label{eq21b}
\end{equation}
We then speculate that $\beta$ becomes independent of $\beta_0$, as $\beta_0$ is increased, as soon as the
postshock magnetic pressure becomes important, because if it were not important there would not be a significant
alignement of flow velocity and magnetic field in regions of compression. Based on the simple approximations leading
to equation (\ref{eq_betamodel}), the postshock magnetic pressure is of the order of the postshock thermal pressure,
or larger, if  ${\cal M}_{\rm A,0}\gtrsim \sqrt{2}\, \beta_0$. This condition is satisfied by the three simulations of \citet{Kritsuk+09}.

In summary, we make the ansatz that the critical density and the standard deviation of the density pdf are given by the
equations (\ref{eq_rhocrmhd}) and (\ref{eq21}) respectively, where $\beta \approx 0.39$ if 
${\cal M}_{\rm A,0}\gtrsim \sqrt{2}\, \beta_0$, which covers all reasonable values of magnetic field strengths and 
Mach numbers in molecular clouds. If ${\cal M}_{\rm A,0}< \sqrt{2}\, \beta_0$, then $\beta \to \infty$ as $\beta_0 \to \infty$,
and both equations reduce to their corresponding non-magnetized forms, given by equations (\ref{eq_rhocr_hd}) and
(\ref{eq18}) respectively.

\section{Star Formation Rate}

In \citet{Padoan+Nordlund04bd} we computed the mass fraction available to form brown dwarfs
as the integral of the pdf of gas density from a critical density to infinity. In that case the critical
density was defined as the density of a critical Bonnor-Ebert sphere with a mass of  0.075~M$_{\odot}$.
\citet{Krumholz+McKee05sfr} used the same integral to compute the total mass available for
star formation, and defined the critical density based on the condition of turbulent support
of a Bonnor-Ebert sphere. Here we follow the same procedure, with the critical density
defined by the condition for magnetic and thermal support expressed by equation (\ref{eq_condition}).

Assuming that a fraction $\epsilon$ of the mass fraction above the critical density is turned into stars in a free-fall time of the
critical density, $\tau_{\rm ff,cr}=(3 \pi/(32 G \rho_{\rm cr,MHD}))^{1/2}$, the star formation rate per free-fall time (the
mass fraction turned into stars in a free-fall time) is given by\footnote{
The integral in equation (\ref{eq22}) is solved assuming that the critical density is not strongly correlated with the local value
of the density, or, equivalently, that the actual postshock thickness is not strongly correlated with the postshock density.
}:
\begin{eqnarray}
{\rm SFR}_{\rm ff} & = & \epsilon \, {\tau_{\rm ff,0}\over \tau_{\rm ff,cr}} \int_{x_{\rm cr}}^\infty x \,p_{\rm MHD}(x)\, dx  \nonumber \\
                                & = & \epsilon \, \frac{x_{\rm cr}^{1/2}}{2}\left(1+{\rm erf} \left[  \frac{\sigma^2-2\,{\rm ln}\left(x_{\rm cr}\right) }
                                {2^{3/2}\,\sigma}\right]\right)
\label{eq22}
\end{eqnarray}
where $\tau_{\rm ff,0}=(3 \pi/(32 G \rho_0))^{1/2}$ is the free-fall time of the mean density,
$x_{\rm cr}=\rho_{\rm cr,MHD}/\rho_0$ given by equation (\ref{eq_rhocrmhd}), $\sigma=\sigma_{\rm MHD}$
given by equation (\ref{eq_sigma_mhd}), and the expression is valid also in the limit of $\beta \to \infty$.

\citet{Krumholz+Tan07slowsf} have argued that the value of SFR$_{\rm ff}$ is approximately the same in
very different star forming environments. If so, the choice of expressing the SFR with a time unit equal to the
free-fall time, introduced in \citet{Krumholz+McKee05sfr}, is useful when comparing with observational
estimates of the SFR. However, if star forming clouds on all scales were mostly transient structures, surviving
only a few local dynamical times in the turbulent flow that formed them, observational estimates of the star
formation efficiency  (rather than the SFR) could be compared directly with the predicted SFR per dynamical time,
${\rm SFR}_{\rm dyn}={\rm SFR}_{\rm ff}\, \tau_{\rm dyn}/\tau_{\rm ff,0}$, where $\tau_{\rm dyn}=R/\sigma_{\rm v,3D}$,
and $R$ is the cloud radius. With this definition of the dynamical time as a crossing time, SFR$_{\rm dyn}$ decreases
with increasing $\alpha_{\rm vir}$ faster than SFR$_{\rm ff}$,
SFR$_{\rm dyn} \propto \alpha_{\rm vir}^{-1/2}$SFR$_{\rm ff}$. \citet{Elmegreen07time} has criticized
the evidence presented by \citet{Krumholz+Tan07slowsf}, in support of his previous suggestion that
the process of star formation lasts approximately 1--2 dynamical times on all scales \citep{Elmegreen00crossing}.
However, he defines the dynamical time as $1/(G\rho)^{1/2}=0.54 \tau_{\rm ff,0}$, assuming that the cloud internal
velocity dispersion is of the order of the virial velocity, which implies SFR$_{\rm dyn}=0.54\,$SFR$_{\rm ff}$.

\begin{figure}[t]
\includegraphics[width=\columnwidth]{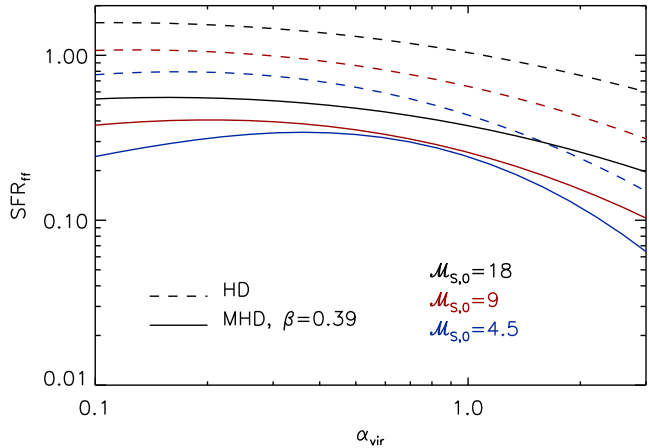}
\caption[]{The star formation rate per free-fall time versus the virial parameter according to equation
(\ref{eq22}), for the HD case (dashed lines), and the MHD case with $\beta=0.39$ (solid lines).
In both cases, the three lines are for three different
values of the sonic rms Mach number, ${\cal M}_{\rm S,0}=4.5$, 9, and 18. In the MHD case, SFR$_{\rm ff}$ has been divided
by a factor of two, in order to separate the MHD curves from the HD ones, and also because the star-formation simulations
show that only half of the mass above the critical density can collapse in the MHD case, as shown in \S7.}
\label{f3}
\end{figure}

In \citet{Krumholz+McKee05sfr}, the timescale for the collapse of the mass above the critical
density is assumed to be proportional to $\tau_{\rm ff,0}$, with the constant of proportionality ($\phi_{\rm t}$ in their
equation (19)) to be determined by comparison with numerical simulations. Our choice of a timescale equal to
$\tau_{\rm ff,cr}$ is physically motivated, because structures of density equal to $\rho_{\rm cr}$ should
collapse on that timescale. Because of our definite and physically motivated choice of the timescale, the prediction
of our model with $\epsilon=1$ should be interpreted as the maximum allowed star formation rate. The value of
SFR$_{\rm ff}$ in the simulations should never be larger than that. If it is smaller than the maximum rate predicted by
the model, the reduction is absorbed by the efficiency factor $\epsilon$, meaning that only a mass fraction $\epsilon$
of the gas with density above the critical one is found within gravitationally unstable regions.

In \S7, we show that the star formation rate in our HD simulations achieves this predicted maximum value, for any
value of $\alpha_{\rm vir}$ we have tested ($\epsilon=1$, independent of $\alpha_{\rm vir}$), while in the MHD
simulations only approximately half of the magnetized gas above the critical density seems to be in collapsing
regions ($\epsilon=0.5$, independent of $\alpha_{\rm vir}$). Because the simulations are reproduced by the
model with $\epsilon$ independent of $\alpha_{\rm vir}$, the timescale $\tau_{\rm ff}$ in \citet{Krumholz+McKee05sfr}
is not a good choice, as it would require their coefficient $\phi_{\rm t}$ to vary with $\alpha_{\rm vir}$ (the
relation $\tau_{\rm ff,0}/ \tau_{\rm ff,cr}=x_{\rm cr}^{1/2}\sim \alpha_{\rm vir}^{1/2}$, shows that our model predicts
a shallower dependence of SFR$_{\rm ff}$ on $\alpha_{\rm vir}$ than  in \citet{Krumholz+McKee05sfr})

Figure~\ref{f3} shows the result of equation (\ref{eq22}) as a function of the virial parameter,
for three values of the sonic Mach number, ${\cal M}_{\rm S,0}=4.5$, 9, and 18, in the MHD case,
$\beta=0.39$, and in the HD case ($\beta=\infty$). We have assumed a value of $\theta=0.35$, as discussed in \S2.
In the HD case (dashed lines) we have assumed $\epsilon=1$, while the curves for the MHD case (solid lines)
are computed for $\epsilon=0.5$. This choice of $\epsilon$ is motivated by the numerical results presented in
\S7.

%
%

Due to our timescale choice of $\tau_{\rm ff,cr}$ instead of $\tau_{\rm ff,0}$, the SFR$_{\rm ff}$ is found to increase with
increasing ${\cal M}_{\rm S,0}$ (for constant virial parameter), while in \citet{Krumholz+McKee05sfr} it decreases with increasing ${\cal M}_{\rm S,0}$,
as shown by their Figure~3 and by the power-law approximation, SFR$_{\rm ff}\sim{\cal M}_{\rm S,0}^{-0.32}$, in their
equation (30). We will show in \S7 that the Mach number dependence of our model is confirmed by the
star-formation simulations.


\section{SFR in Simulations of Driven MHD Turbulence}

In order to test the SFR model, we have run a set of simulations of driven supersonic turbulence, on meshes
with $500^3$- $1,000^3$ computational zones. Using the same methods and setup as in
\citet{Padoan+Nordlund02imf} and \citet{Padoan+Nordlund04bd}, we adopt periodic boundary conditions,
isothermal equation of state, and random forcing in Fourier space at wavenumbers $1\le k\le 2$ ($k=1$
corresponds to the computational box size). The simulations are all based on two initial
snapshots of fully developed turbulence, one for HD and one for MHD. These snapshots are obtained
by running the HD and the MHD simulations from initial states with uniform initial
density and magnetic field, and random initial
velocity field with power only at wavenumbers $1 \le k\le 2$, for approximately 5 dynamical times, on meshes with $1,000^3$
computational zones, with the driving force keeping the rms sonic Mach number at the approximate value of
${\cal M}_{\rm S,0}=\sigma_{\rm v,3D}/c_{\rm S} \approx 9$. In the case of the HD run with ${\cal M}_{\rm S,0}=4.5$
(run HD10 in Table~\ref{t1}), the forcing was reduced prior to the inclusion of self-gravity until the targeted
Mach number was reached.

In the MHD simulation, the initial magnetic field is such that the initial value of the ratio of gas to
magnetic pressure is $\beta_0=22.2$. At the time when the gravitational force is included, the magnetic
field has been amplified by the turbulence, and the value of $\beta$ is
$\beta = 2 \, c_{\rm S}^2 / \langle B^2 / 4\pi \rho \rangle = 0.33$, consistent with equation (\ref{eq_betamodel}) with $b=0.63$,
using the mean squared $v_{\rm A}$ averaged in regions with density larger than twice the mean, and $\beta=0.11$,
consistent with equation (\ref{eq_betamodel}) with $b=0.22$, using the mean squared $v_{\rm A}$ averaged over the whole domain.

The star formation simulations start when the gravitational force is included.
The computational mesh is downsized from
$1,000^3$ to $500^3$ zones for the $500^3$ runs, or kept the same for the $1,000^3$ runs.
The driving force is still active during the star-formation phase of the simulations, in order to achieve a
stationary value of $\alpha_{\rm vir}$ to correlate with the SFR.

\begin{figure}[t]
\includegraphics[width=\columnwidth]{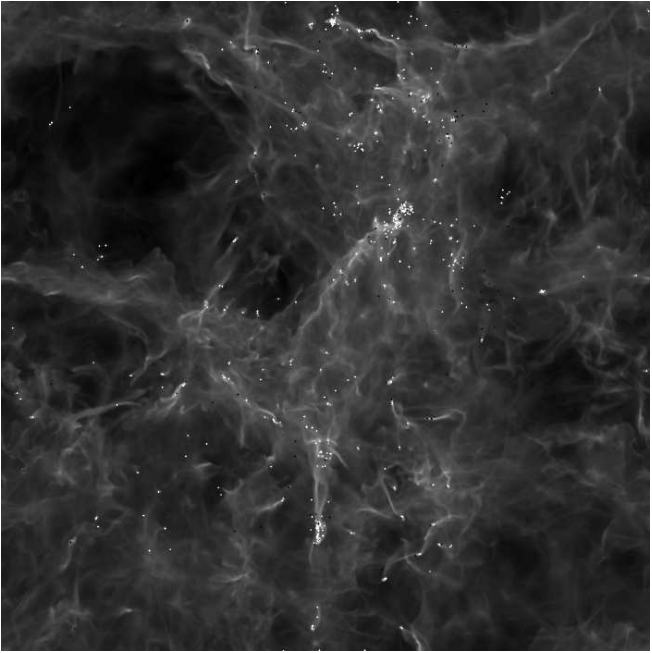}
\caption[]{Logarithm of projected density from a snapshot of an exploratory $1,000^3$ run with $\beta_0=22.2$,
${\cal M}_{\rm S,0}=18$, and $\alpha_{\rm vir}=0.9$, at a time when approximately 10\% of the mass has been converted
into stars. Bright dots show the positions of the stars (sink particles), while black dots are for brown dwarfs (some of which
are still accreting and may later grow to stellar masses). 
}
\label{f5}
\end{figure}

An example of a projected density field from a star formation simulation is shown in Figure~\ref{f5}.
Table~1 gives the values of the sonic rms Mach number,  ${\cal M}_{\rm S,0}$, the initial pressure ratio, $\beta_0$,
the Jeans length in units of the box size, $L_{\rm J}/L_0 \approx 1.94 \,\alpha_{\rm vir}^{1/2} {\cal M}_{\rm S,0}^{-1}$,
the virial parameter, $\alpha_{\rm vir}$, and the SFR per free-fall time, SFR$_{\rm ff}$, for all the 19 simulations used to
test the theoretical model.

Our simulations represent an intermediate range of scales. The forcing represents the inertial forcing from
scales larger than the box size.  These larger scale motions have longer turn-over times -- and hence longer
life times -- than the turn-over times of the scales covered by the simulations. They act to maintain the kinetic
energy on smaller scales.  Without the corresponding driving, the motions on the scales covered by the
simulations would decay, which would lead to a lowering of the virial parameter and a corresponding secular
increase in the star formation rate.  By maintaining the driving we avoid the secular evolution and obtain
a consistent and nearly constant star formation rate.

The virial parameter defined in equation (\ref{alpha}) is for a sphere of uniform density.
The simulations are carried out in a cubic domain and generate a highly nonlinear density field;
real star forming regions have irregular shapes and are highly fragmented. The virial parameter of the
simulations, as well as that of real molecular clouds, is therefore only an approximation of the energy ratio.
To define the virial parameter of the simulations, we have chosen to use equation (\ref{alpha}), with
$R=L_0/2$, where $L_0$ is the box size, and $M$ equal to the total mass in the box, $M_0$.
The virial parameter is then $\alpha_{\rm vir}= 5 \, v_0^2 \, L_0 / (6 \, G M_0)$, where $v_0$ is the
three-dimensional rms velocity in the box.

\begin{table}[t]
\caption{Non-dimensional parameters of the simulations used to measure the star formation rate.}
\centering
\begin{tabular}{llccccccc}
\hline\hline \\[-2.2ex]
Run & \,\,\, $N$ \,\,\, & ${\cal M}_{\rm S,0}$ & \,\, $\beta_0$ \,\, & \,$L_{\rm J}/L_0$  & \,\, $\alpha_{\rm vir}$ \,\, &  SFR$_{\rm ff}$
\\ [0.8ex]
\hline \\[-1.8ex]
HD1       & $500^3$     & 9.0 &   $\infty$      &   0.10                   &   0.22                        &  1.01                      \\
HD2       & $500^3$     & 9.0 &   $\infty$      &   0.12                   &   0.34                        &  0.86                      \\
HD3       & $500^3$     & 9.0 &   $\infty$      &   0.15                   &   0.48                        &  0.86                      \\
HD4       & $500^3$     & 9.0 &   $\infty$      &   0.18                   &   0.67                        &  0.75                      \\
HD5       & $500^3$     & 9.0 &   $\infty$      &   0.21                   &   0.95                        &  0.68                      \\
HD6       & $500^3$     & 9.0 &   $\infty$      &   0.25                   &   1.33                        &  0.51                      \\
HD7       & $500^3$     & 9.0 &   $\infty$      &   0.31                   &   2.04                        &  0.29                      \\
HD8       & $1000^3$   & 9.0 &   $\infty$      &   0.21                   &   0.95                        &  0.71                      \\
HD9       & $1000^3$   & 9.0 &   $\infty$      &   0.31                   &   2.04                        &  0.41                      \\
HD10     & $500^3$     & 4.5 &   $\infty$      &   0.18                   &   0.67                        &  0.54                      \\
MHD1    & $500^3$     & 9.0 &     22.2         &   0.10                   &   0.22                        &  0.43                      \\
MHD2    & $500^3$     & 9.0 &     22.2         &   0.12                   &   0.34                        &  0.42                      \\
MHD3    & $500^3$     & 9.0 &     22.2         &   0.15                   &   0.48                        &  0.39                      \\
MHD4    & $500^3$     & 9.0 &     22.2         &   0.18                   &   0.67                        &  0.31                      \\
MHD5    & $500^3$     & 9.0 &     22.2         &   0.21                   &   0.95                        &  0.19                      \\
MHD6    & $500^3$     & 9.0  &    22.2         &   0.25                   &   1.33                        &  0.15                      \\
MHD7    & $500^3$     & 9.0  &    22.2         &   0.31                   &   2.04                        &  0.05                      \\
MHD8    & $1000^3$   & 9.0  &    22.2         &   0.21                   &   0.95                        &  0.20                      \\
MHD9    & $1000^3$   & 9.0  &    22.2         &   0.31                   &   2.04                        &  0.15
\\ [0.4ex]
\hline \\
\end{tabular}
\label{t1}
\end{table}

A collapsing region is captured by the creation of an accreting sink particle if the density exceeds a
certain density threshold (8,000 times the mean density in both $500^3$ and $1000^3$ runs).
We have verified that the largest density reached by non-collapsing regions is
always much smaller than that value, so only a collapsing region can create a sink particle. No other
conditions need to be satisfied to identify genuine collapsing regions. Once a particle is created its
subsequent motion is followed, allowing for influences from the gravitational potential
and from accretion.  When calculating the gravitational potential the masses of the stars are added back
into a fiducial density field, using narrow Gaussian profiles ($1/e$ radius 1.15 grid zones) to represent the sink particles.
The Poisson equation for the gravitational potential is solved using parallelized Fast Fourier
Transforms with Gaussian softening ($1/e$ radius $2\sqrt{2}$ grid zones).

Further accretion (defined as density exceeding the density threshold) is collected onto the nearest
sink particle if the distance is less than four grid zones.  Sink particles are not merged, and
thus maintain their identity even if they become trapped in the same potential well (the softening
of the gravitational potential ensures that no singularity occurs).

\begin{figure}[t]
\includegraphics[width=\columnwidth]{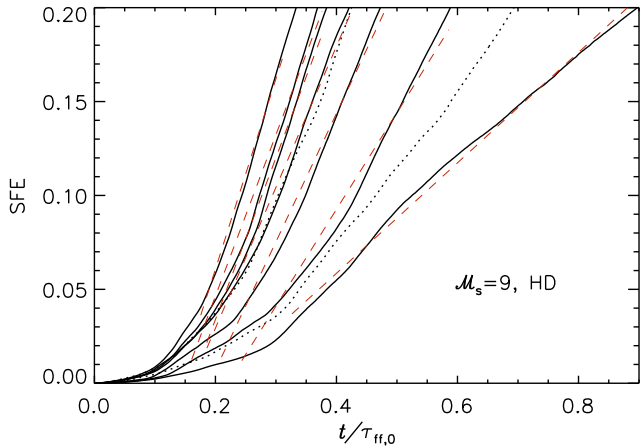}
\caption[]{Star-formation efficiency versus time for all the the HD runs with ${\cal M}_{\rm S,0}= 9$ listed in Table~\ref{t1}.
The star formation efficiency is defined as the mass in stars (sink particles) divided by the total mass, and the time is in
units of the free-fall time of the mean density of each simulation, $\tau_{\rm ff,0}$. The dashed lines show the least-squares
fit to each curve between SFE=0.03 and SFE=0.2. The slope of those linear fits defines the SFR$_{\rm ff}$ plotted in
Figure~\ref{f8}. The two dotted curves are from the $1,000^3$ runs (HD8 and HD9). The value of $\alpha_{\rm vir}$ for
each curve varies from 0.22 to 2.04 (see Table~\ref{t1}) from top to bottom. }
\label{f6}
\end{figure}

\begin{figure}[t]
\includegraphics[width=\columnwidth]{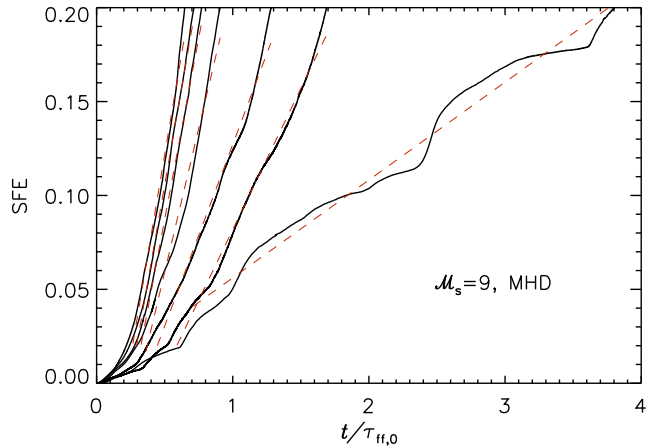}
\caption[]{Same as Figure~\ref{f6}, but for the MHD simulations.}
\label{f7}
\end{figure}

In Figures~\ref{f6} and \ref{f7} we show the star formation efficiency (SFE) versus time in the HD and MHD
simulations respectively. The  SFE is defined as the mass in sink particles divided by the total initial
mass. The time is given in units of the free-fall time of each simulation, so the slope of the plots
corresponds to the SFR$_{\rm ff}$. The plots only show the SFE from the time when the first sink particle
is created, which is some time after the gravity is turned on. The time to the formation of the first sink
particles is longer for simulations with larger $\alpha_{\rm vir}$, which cannot be appreciated in 
Figures~\ref{f6} and \ref{f7}.

Even after the first sink particle
is created, there is still an initial transient phase with increasing SFR. This transient phase usually
lasts until SFE$\approx 0.03$. The SFR$_{\rm ff}$ is therefore estimated as the slope of a least-square
fit to the SFE in the interval 0.03$\le$SFE$\le$0.2. The SFR based on this interval of the SFE is quite
robust with respect to changes in the treatment of sink particles (threshold density, accretion radius,
gravitational softening, etc.); even changes that affect the number of sink particles significantly do
not change the measured SFR much.

All runs were continued until SFE$\ge$0.4, and some
until SFE$\approx$0.8. However, we prefer to fit the SFR only up to SFE=0.2, because larger values are
rarely found in star forming regions, and because the simulations should not be trusted past that point (the stellar
content may start to affect the gas motion, and the gravitational interactions in close encounters between sink particles
are not accurately computed with an N-body code).

Figures~\ref{f6} and \ref{f7} show that SFR$_{\rm ff}$ decreases monotonically with increasing $\alpha_{\rm vir}$
($\alpha_{\rm vir}=0.22$ to 2.04 for the plots from top to bottom). It is well defined because
the SFE plots are almost straight lines (constant instantaneous SFR) for almost all the simulations, except for a
tendency of some of the MHD runs to slightly increase their SFR also at relatively high values of SFE.
The MHD run with the highest virial parameter, $\alpha_{\rm vir}=2.04$ (MHD7 in Table~\ref{t1}),
has the lowest SFR and shows a rather episodic SFE evolution. However, its mean SFR$_{\rm ff}$ in the range
0.03$\le$SFE$\le$0.2 is well defined. \\

\section{Models versus Numerical Results}

\begin{figure}[t]
\includegraphics[width=\columnwidth]{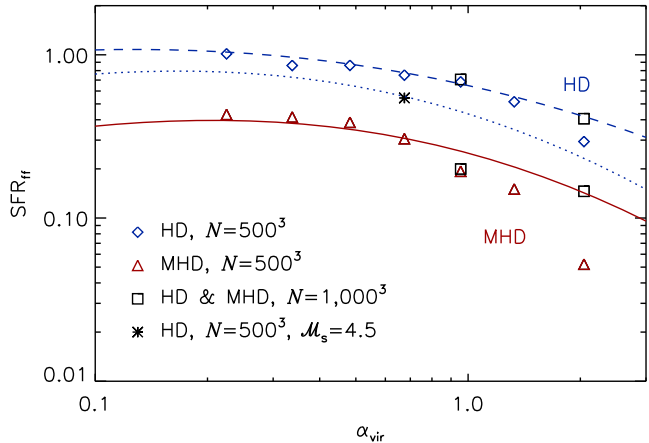}
\caption[]{Star formation rate per free-fall time versus virial parameter for the $500^3$ MHD simulations (triangles)
and for the $500^3$ HD simulations (diamonds) with ${\cal M}_{\rm S,0}= 9$. The squares are for the $1,000^3$ runs,
and the asterisk for the $500^3$ HD run with ${\cal M}_{\rm S,0}= 4.5$. The MHD model with ${\cal M}_{\rm S,0}= 9$
and $\beta=0.39$ is shown by the solid line. 
The HD model ($\beta_0=\infty$) is shown by the dashed line for ${\cal M}_{\rm S,0}= 9$, and by the
dotted line for ${\cal M}_{\rm S,0}= 4.5$. The values of SFR$_{\rm ff}$ from the simulations are the slopes of the linear
fits to the SFE versus time between SFE=0.03 and SFE=0.2 (see Figures~\ref{f6} and \ref{f7}). The values of
SFR$_{\rm ff}$ of the models are from equation (\ref{eq22}), with $\epsilon=1.0$ in the HD case, and $\epsilon=0.5$
for the MHD model. \\
}
\label{f8}
\end{figure}

Figure~\ref{f8} compares the SFR model with the numerical results. It shows SFR$_{\rm ff}$ versus
$\alpha_{\rm vir}$ for all the simulations listed in Table~1, and for the HD (dashed and dotted lines) and
MHD (solid line) models. The model prediction corresponds to equation (\ref{eq22}), with $\epsilon=1.0$
in the HD case, and $\epsilon=0.5$, in the MHD case.

The HD simulations follow almost exactly the theoretical prediction with $\epsilon=1$, suggesting that all the gas
with density above the critical value collapses in a timescale of order $\tau_{\rm ff,cr}$, as assumed in the model.
The dependence of SFR$_{\rm ff}$ on $\alpha_{\rm vir}$ is too shallow to be consistent with the parametrization
in \citet{Krumholz+McKee05sfr}, where the timescale is $\phi_{\rm t} \tau_{\rm ff,0}$, unless $\phi_{\rm t}$ is
allowed to change with the virial parameter, $\phi_{\rm t}\propto \alpha_{\rm vir}^{1/2}$.

Of the $500^3$ HD runs, only the one with the highest $\alpha_{\rm vir}$ (HD7) deviates significantly ($\approx$50\%)
from the theoretical prediction. However, the corresponding higher resolution run yields a higher value of SFR$_{\rm ff}$,
nearly identical to the theoretical prediction. At $\alpha_{\rm vir}=0.95$, instead, the $500^3$ run is already converged to
the SFR of the corresponding $1,000^3$ run (HD5 and HD8 respectively). The run with the highest value of $\alpha_{\rm vir}$
is expected to be the one requiring the largest numerical resolution, because $\rho_{\rm cr,HD}/ \rho_0 \propto \alpha_{\rm vir}$,
according to equation (\ref{eq_rhocr_hd}). In other words, higher $\alpha_{\rm vir}$ can be interpreted as lower mean density (everything else remaining unchanged), making it harder to reach the critical density for collapse in the simulation.

The HD runs also confirm the theoretical prediction that SFR$_{\rm ff}$ should increase with increasing ${\cal M}_{\rm S,0}$
(the opposite of the prediction in \citet{Krumholz+McKee05sfr}), as shown by the comparison of the runs HD4 and HD10,
with ${\cal M}_{\rm S,0}=4.5$ and 9, respectively. The lower Mach number run fits very well the theoretical prediction,
confirming our choice of $\tau_{\rm ff,cr}$ for the timescale of star formation.

Similar considerations apply to the MHD runs. There is good agreement between the simulations and the theoretical
model with $\epsilon=0.5$, although the model predicts a significantly higher SFR than the $500^3$ simulation with the largest value 
of $\alpha_{\rm vir}$. 
This discrepancy may be entirely due to the insufficient numerical rsolution of the simulation, because the $1,000^3$ run with
the same virial parameter, $\alpha_{\rm vir}=2.04$, yields a value of SFR$_{\rm ff}$ almost identical to the theoretical prediction.
Like in the HD simulations, the case with $\alpha_{\rm vir}=0.95$ seems to be already converged at a resolution of $500^3$
computational zones, as its SFR$_{\rm ff}$ is nearly identical to that of the corresponding $1,000^3$ run (and only approximately
20\% below the predicted value).

The value of $\epsilon=0.5$ derived from the comparison of the model with the MHD simulations can be understood with the
following argument. In the MHD case, the critical mass for collapse depends on both the local density and the local magnetic
field strength, $B$, while in HD it depends only on the local density (assuming constant temperature). In HD, regions with
density larger than the critical value must collapse, because of the lack of gas pressure support. In the MHD case, instead,
at any value of density there is a large scatter in $B$. Even if the local density is above the critical value, the pressure support
is dominated by magnetic pressure, and, due to the large scatter in $B$ (and even larger in $B^2$), a region with $B$ larger
than the mean value at that density may be prevented from collapsing. The value $\epsilon=0.5$ is reasonable, because
the critical density is derived with characteristic postshock values, and it is possible that in half of the cases the magnetic field
deviates enough from its characteristic postshock value to prevent the gravitational collapse.

\section{Discussion}

\subsection{The Timescale of Star Formation}

We have modeled the SFR assuming that a mass fraction $\epsilon$ of all regions with density larger than the critical one
is converted into stars in a time $\tau_{\rm ff,cr}$. To maintain this SFR over a time longer than $\tau_{\rm ff,cr}$, the
turbulent flow must continuously ``regenerate'' the high density tail of the gas density pdf, on a timescale shorter than or
equal to $\tau_{\rm ff,cr}$. This may seem unlikely, because in super-Alfv\'{e}nic turbulence the global dynamical time is
always longer than the collapse time of the unstable regions, $\tau_{\rm dyn} > \tau_{\rm ff,cr}$. For example,
$\tau_{\rm ff,cr} / \tau_{\rm dyn}= 4.6 \, \theta \, {\cal M}^{-1}_{\rm S,0}(1+\beta^{-1})/(1+0.925 \beta^{-3/2})^{1/3}=0.34$
and 0.16, for $\beta=0.39$ and $\beta\to \infty$ respectively, assuming ${\cal M}_{\rm S,0}=10$. However,
the turbulence can ``regenerate'' the high density tail of the pdf sufficiently rapidly, because the collapsing dense regions
account for only a very small fraction of the total mass. The turbulent flow ``processes'' a gas mass of the order of the
total mass in one dynamical time (think of the trivial example of a single shock crossing the whole volume in one dynamical
time), hence a mass fraction of order $\tau_{\rm ff,cr} / \tau_{\rm dyn}$ in a time equal to $\tau_{\rm ff,cr}$. At a characteristic
Mach number value of ${\cal M}_{\rm S,0}=10$, this mass fraction is always larger than 0.16 (the value found above for the
extreme limit of $\beta \to \infty$). The mass fraction above the critical density is typically much smaller than that, of order
a few percent. If the critical density is increased, the mass fraction processed by the turbulence in a time $\tau_{\rm ff,cr}$
decreases like $\rho_{\rm cr}^{-1/2}$, while the mass fraction above the critical density drops more rapidly, due to the
Log-Normal nature of the pdf. Therefore, the collapse of regions with density above the critical value can be continuously
``fed'' by the turbulence, and our choice of $\tau_{\rm ff,cr}$ as the star formation timescale is justified.

The above argument also means that the collapse of unstable regions of densities $\rho > \rho_{\rm cr}$ is not expected to
strongly affect the density pdf at densities $\rho \le \rho_{\rm cr}$. We have verified that, once star formation is initiated by the
inclusion of self-gravity, the density pdf in our simulations develops a power law tail $\propto \rho^{-3/2}$ at densities
$\rho \gtrsim \rho_{\rm cr}$, a signature of free-fall, while it maintains the Log-Normal shape for $\rho < \rho_{\rm cr}$.
The rapid mass processing by the turbulence that allows the preservation of the Log-Normal pdf despite the effect of self-gravity explains why it is possible to predict the SFR based on the statistics of turbulence alone, with no modification due to self-gravity.
One can model the process of star formation with two distinct phases: the formation of dense regions
by turbulent compressions, and the gravitational collapse of the densest of those regions. Locally, this is roughly what happens,
while globally, the turbulence and the gravity are always operating at the same time.

\subsection{SFR in Molecular Clouds}

\citet{Krumholz+Tan07slowsf} argue that SFR$_{\rm ff}\sim 0.02$ in a variety of star forming environments,
spanning approximately four orders of magnitude in gas density. The estimated values of SFR have large
error bars, but the lack of a strong density dependence would suggest that most star forming regions
have a comparable value of $\alpha_{\rm vir}$. More recently, \citet{Evans+09} have estimated values of
SFR$_{\rm ff}$ in giant molecular clouds (GMCs) and within some of the dense cloud cores. They find values significantly
larger than the characteristic one in \citet{Krumholz+Tan07slowsf}. They obtain SFR$_{\rm ff}=0.03$ to 0.06
for GMCs with mean densities distributed around a mean value of $\langle n \rangle=390$~cm$^{-3}$ (and SFE
in the range 0.03 to 0.06 as well), and SFR$_{\rm ff}=0.05$ to 0.25 for dense cores with mean densities 50-200 times those
of the GMCs (and SFE of approximately 0.5). These values are computed by assuming that all the stars detected
(by their infrared excess) have been formed in the last 2~Myr. The authors report a best estimate of 2$\pm$1~Myr
for the lifetime of the Class II phase, meaning that the SFR$_{\rm ff}$ could be 50\% lower, or 100\% higher than the
values given above. Accounting for this uncertainty, one gets SFR$_{\rm ff}=0.02$ to 0.12 for GMCs, and
SFR$_{\rm ff}=0.03$ to 0.5 for dense cores, suggesting a characteristic value of order 0.1, rather than 0.01.
\citet{Evans+09} suggest that the SFR in dense cores would be lower, if one assumed that the total mass in the
cores was larger when the star formation process started than at present. However, it is also possible that star
formation was already occurring while the cores were still being assembled by flows accreting from the larger
scale. In this case, the initial core mass may have been smaller than the current one, resulting in a SFR larger
than estimated.

\begin{figure}[t]
\includegraphics[width=\columnwidth]{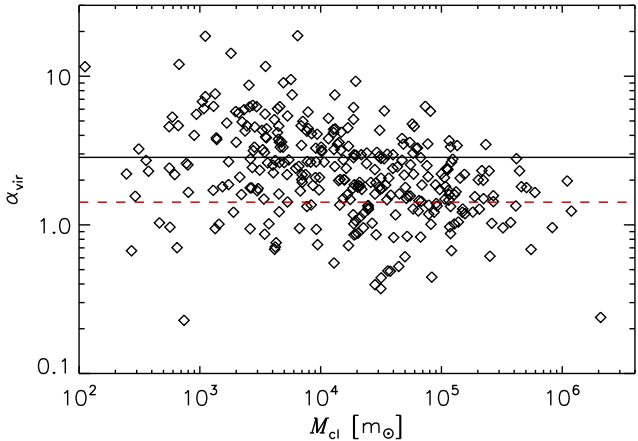}
\caption[]{Virial parameter versus cloud mass from \citet{Heyer+09virial}. All 316 objects from their Table~1 are shown, including
those selected within the smaller maps of area $A_2$. The horizontal solid line marks the mean value of $\alpha_{\rm vir}=2.8$,
and the dashed line half that value, $\alpha_{\rm vir}=1.4$, assuming that the LTE-derived mass underestimates the true mass
by a factor of two, causing an overestimates of the virial parameter by the same factor. \\
}
\label{f9}
\end{figure}

Figure~\ref{f3} shows that for a range of values of ${\cal M}_{\rm S,0}$ characteristic 
of MCs, we predict SFR$_{\rm ff}\approx 0.12$ to 0.28 at $\alpha_{\rm vir}=2$. These values should be reduced by a factor 
of two or three \citep{Matzner+McKee2000,Andre+2010}, to account for mass loss from stellar
outflows and jets, not included in the model and in the simulations. With this reduction, our results are consistent with 
the relatively high values of SFR$_{\rm ff}$ found by \citet{Evans+09}. Another source of uncertainty lies in the
mapping of our definition of the virial ratio for a periodic box (Eq.\ \ref{alpha}) to the virial ratios used
to characterize observed star forming regions.  One example is the estimate of the characteristic 
$\alpha_{\rm vir}$ of GMCs. \citet{Heyer+09virial} have recently studied again a large
subset of the GMCs sample of  \citet{Solomon+87}. For each cloud, they compute LTE masses based on the J=1-0
emission lines of $^{13}$CO and $^{12}$CO. They find masses smaller by a factor of 2 to 5 than the virial masses
derived by \citet{Solomon+87}. Their revised velocity dispersion are also somewhat smaller than in \citet{Solomon+87},
but their resulting virial parameters are still a factor of approximately 2-3 larger. 

Figure~\ref{f9} shows
$\alpha_{\rm vir}$ versus the cloud mass, $M_{\rm cl}$, for all their 316 maps (including the smaller ones of area $A_2$).
The mean value of the virial parameter is $\alpha_{\rm vir}= 2.8\pm2.4$. If the LTE-derived mass underestimates the real
mass by a factor up to two, as argued by the authors, then the values of $\alpha_{\rm vir}$ should be reduced by a factor
of two. The mean value is therefore likely to lie in the range $\alpha_{\rm vir} = 1.4$ to 2.8, but with a very large scatter.
As commented above, if GMCs have a characteristic value of $\alpha_{\rm vir}\approx 2$, as suggested by this
observational sample, the SFR predicted by our model for a reasonable range of values of ${\cal M}_{\rm S,0}$, and
accounting for a factor of two or three reduction due to mass-loss in outflows and jets, may be consistent with the recent 
observational estimates by \citet{Evans+09}.

\section{Summary and Concluding Remarks}

This work presents a new physical model of the SFR that could be implemented in galaxy formation simulations.
The model depends on the relative importance of gravitational, turbulent, magnetic, and thermal energies,
expressed through the virial parameter, $\alpha_{\rm vir}$, the rms sonic Mach number, ${\cal M}_{\rm S,0}$,
and the ratio of the mean gas pressure to mean magnetic pressure, $\beta_0$. The value of SFR$_{\rm ff}$ is predicted
to decrease with increasing $\alpha_{\rm vir}$, and to increase with increasing ${\cal M}_{\rm S,0}$,
for values typical of star forming regions (${\cal M}_{\rm S,0}\approx 4$-20).  In the
complete absence of a magnetic field, SFR$_{\rm ff}$ increases typically by a factor of three, proving the
importance of magnetic fields in star formation, even when they are relatively weak (super-Alfv\'{e}nic turbulence).

In the non-magnetized limit, our definition of the critical density for star formation has the same dependence on
$\alpha_{\rm vir}$ and ${\cal M}_{\rm S,0}$ as in the model of \citet{Krumholz+McKee05sfr}, but our physical derivation
does not rely on the concepts of local turbulent pressure and sonic scale. Due to our different choice of star formation 
timescale (see \S~8.1), our model predicts a different dependence of the SFR on $\alpha_{\rm vir}$ and ${\cal M}_{\rm S,0}$ 
than the model of \citet{Krumholz+McKee05sfr}.
The model predictions have been tested with an unprecedented set of large numerical simulations of supersonic MHD
turbulence, including the effect of self-gravity, and capturing collapsing cores as accreting sink particles. The SFR in the
simulations follow closely the theoretical predictions.

Although based on reasonable physical assumptions, this phenomenological model of the SFR bypasses
the great complexity of the nonlinear dynamics of supersonic, self-gravitating, magnetized turbulence, by taking
advantage of the gas density pdf of fully developed turbulence. Because it provides a
prediction of the SFR based solely on turbulence statistics, with no correction for the effect of
self-gravity, the model shows that the process of star formation may be envisioned as the effect of two almost
independent steps: i)  turbulent fragmentation, with little influence from self-gravity, and ii) the
local collapse of the densest regions, with little influence from turbulence. This approximation is a basic
assumption in the stellar IMF model of \citet{Padoan+Nordlund02imf} as well. It is also fundamentally
different from the assumptions of star formation models relying on the concept of local turbulent pressure
support, where the local competition between turbulence and self-gravity is always important on all scales.

This work illustrates how the turbulence controls the SFR. It does not address how the turbulence is driven
to a specific value of $\alpha_{\rm vir}$. Because much of the turbulence driving is likely due to SN
explosions, the turbulent kinetic energy and the value of $\alpha_{\rm vir}$ are coupled to the SFR in
a feedback loop. The feedback determines the equilibrium level of the SFR (and hence also the equilibrium
level of $\alpha_{\rm vir}$) at large scales. If $\alpha_{\rm vir}$ were to decrease (increase) relative
to the equilibrium, the SFR would increase (decrease), according to the results
of this work, resulting in an increased (decreased) energy injection rate by SN explosions, thus restoring a
higher (lower) value of $\alpha_{\rm vir}$. The dependence of the SFR on $\alpha_{\rm vir}$
found in this work suggests that this self-regulation may work quite effectively.

Cosmological simulations of galaxy formation provide the rate of gas cooling and infall, which sets the gas
reservoir for the star formation process and thus ultimately controls the SFR. They also include
prescriptions for the star formation feedback, known to be essential to recover observed properties of
galaxies \citep{Gnedin+09,Gnedin+10}. Future galaxy formation simulations should adopt a physical SFR law with
an explicit dependence on $\alpha_{\rm vir}$, ${\cal M}_{\rm S,0}$, and $\beta$ as derived in this work, in order
to correctly reflect specific conditions of protogalaxies at different redshifts. This requires a treatment of the star
formation feedback capable of providing an estimate of  $\alpha_{\rm vir}$ on scales of order 10-100~pc, not far
from the spatial resolution currently achieved by the largest cosmological simulations of galaxy formation.

\acknowledgements

We are grateful to the anonymous referee and to Liubin Pan, Chris McKee, and Mark Krumholz for reading
the manuscript and providing useful comments. We thank Mark Heyer for providing the data from Table~1 of
\citet{Heyer+09virial} in digital form. This research was supported in part by the NASA ATP grant
NNG056601G, NSF grant AST-0507768, and a grant from the Danish Natural Science Research Council.
This work was prepared in part during the workshop `Star Formation Through Cosmic Time' at the KITP in
Santa Barbara, and was supported in part by the National Science Foundation under Grant No. PHY99-07949.
We utilized computing resources provided by the San Diego Supercomputer Center, by the NASA High End
Computing Program, and by the Danish Center for Scientific Computing.

\bibliographystyle{apj}

\end{document}